\documentclass[prb,amsmath,amssymb,superscriptaddress,twocolumn]{revtex4}
\usepackage{float}
\usepackage{amsmath}
\usepackage{amssymb}
\usepackage{amsfonts}
\usepackage{euscript}
\usepackage{enumerate}
\usepackage{hhline}
\usepackage{pslatex}
\usepackage{tabularx}
\usepackage[usenames,dvipsnames]{xcolor}

\usepackage{graphicx}

\usepackage{dcolumn}
\usepackage{bm}
\usepackage[sort&compress]{natbib}

\usepackage{lipsum}

\newcommand*{\citen}[1]{%
  \begingroup
    \romannumeral-`\x 
    \setcitestyle{numbers}%
    \cite{#1}%
  \endgroup
}

\makeatletter
\renewcommand{\@biblabel}[1]{#1. }
\renewcommand{\@dotsep}{500}
\renewcommand{\@pnumwidth}{0em}
\renewcommand{\l@figure}[2]{
\@dottedtocline{1}{1.5em}{2em}{Figure #1}{}\vspace{15pt}}
\newcommand{\ks}[1]{\textcolor{black}{#1}}
\newcommand{\xl}[1]{\textcolor{black}{#1}}
\usepackage[normalem]{ulem}

\begin{document}

\title{A universal frequency engineering tool for microcavity nonlinear optics: multiple selective mode splitting of whispering-gallery resonances}

\author{Xiyuan Lu}\email{xiyuan.lu@nist.gov}
\affiliation{Microsystems and Nanotechnology Division, Physical Measurement Laboratory, National Institute of Standards and Technology, Gaithersburg, MD 20899, USA}
\affiliation{Institute for Research in Electronics and Applied Physics and Maryland NanoCenter, University of Maryland,
College Park, MD 20742, USA}
\author{Ashutosh Rao}
\affiliation{Microsystems and Nanotechnology Division, Physical Measurement Laboratory, National Institute of Standards and Technology, Gaithersburg, MD 20899, USA}
\affiliation{Department of Chemistry and Biochemistry, University of Maryland,
College Park, MD 20742, USA}
\author{Gregory Moille}
\affiliation{Microsystems and Nanotechnology Division, Physical Measurement Laboratory, National Institute of Standards and Technology, Gaithersburg, MD 20899, USA}
\affiliation{Joint Quantum Institute, NIST/University of Maryland,
College Park, MD 20742, USA}
\author{Daron A. Westly}
\affiliation{Microsystems and Nanotechnology Division, Physical Measurement Laboratory, National Institute of Standards and Technology, Gaithersburg, MD 20899, USA}
\author{Kartik Srinivasan} \email{kartik.srinivasan@nist.gov}
\affiliation{Microsystems and Nanotechnology Division, Physical Measurement Laboratory, National Institute of Standards and Technology, Gaithersburg, MD 20899, USA}
\affiliation{Joint Quantum Institute, NIST/University of Maryland, College Park, MD 20742, USA}

\date{\today}

\begin{abstract}
       {Whispering-gallery microcavities have been used to realize a variety of efficient parametric nonlinear optical processes through the enhanced light-matter interaction brought about by supporting multiple high quality factor and small modal volume resonances. Critical to such studies is the ability to control the relative frequencies of the cavity modes, so that frequency matching is achieved to satisfy energy conservation. Typically this is done by tailoring the resonator cross-section. Doing so modifies the frequencies of all of the cavity modes, that is, the global dispersion profile, which may be undesired, for example, in introducing competing nonlinear processes. Here, we demonstrate a frequency engineering tool, termed multiple selective mode splitting (MSMS), that is independent of the global dispersion and instead allows targeted and independent control of the frequencies of multiple cavity modes. In particular, we show controllable frequency shifts up to 0.8~nm, independent control of the splitting of up to five cavity modes with optical quality factors $\gtrsim$ 10$^5$, and strongly suppressed frequency shifts for untargeted modes. The MSMS technique can be broadly applied to a wide variety of nonlinear optical processes across different material platforms, and can be used to both selectively enhance processes of interest and suppress competing unwanted processes.}
\end{abstract}

\maketitle
\section{Introduction}
Parametric nonlinear optics based on the second-order ($\chi^{(2)}$) or third-order ($\chi^{(3)}$) nonlinear interaction of light has been studied in a variety of optical materials in both non-resonant geometries (bulk crystals, fibers, waveguides) and resonant structures (fabry-perot cavities, millimeter-size cavities, and microcavities). Efficient nonlinear mixing based on these processes requires intense light with frequency and phase matching~\cite{Agrawal2007,Boyd2008}. Among these structures, optical microcavities that confine light in whispering-gallery modes (WGMs) with high quality factors and small mode volumes and therefore enhance light-matter interactions in both space and time have been particularly successful~\cite{Vahala_Nature_2003,Strekalov_2016}. However, such an enhancement requires the interacting WGMs to be frequency matched, that is, their resonance frequencies must satisfy energy conservation with an accuracy approximately given by their linewidths. Realization of frequency matching requires consideration of both the specific material platform and the targeted nonlinear optical process.

\begin{figure*}[t!]
\centering\includegraphics[width=0.90\linewidth]{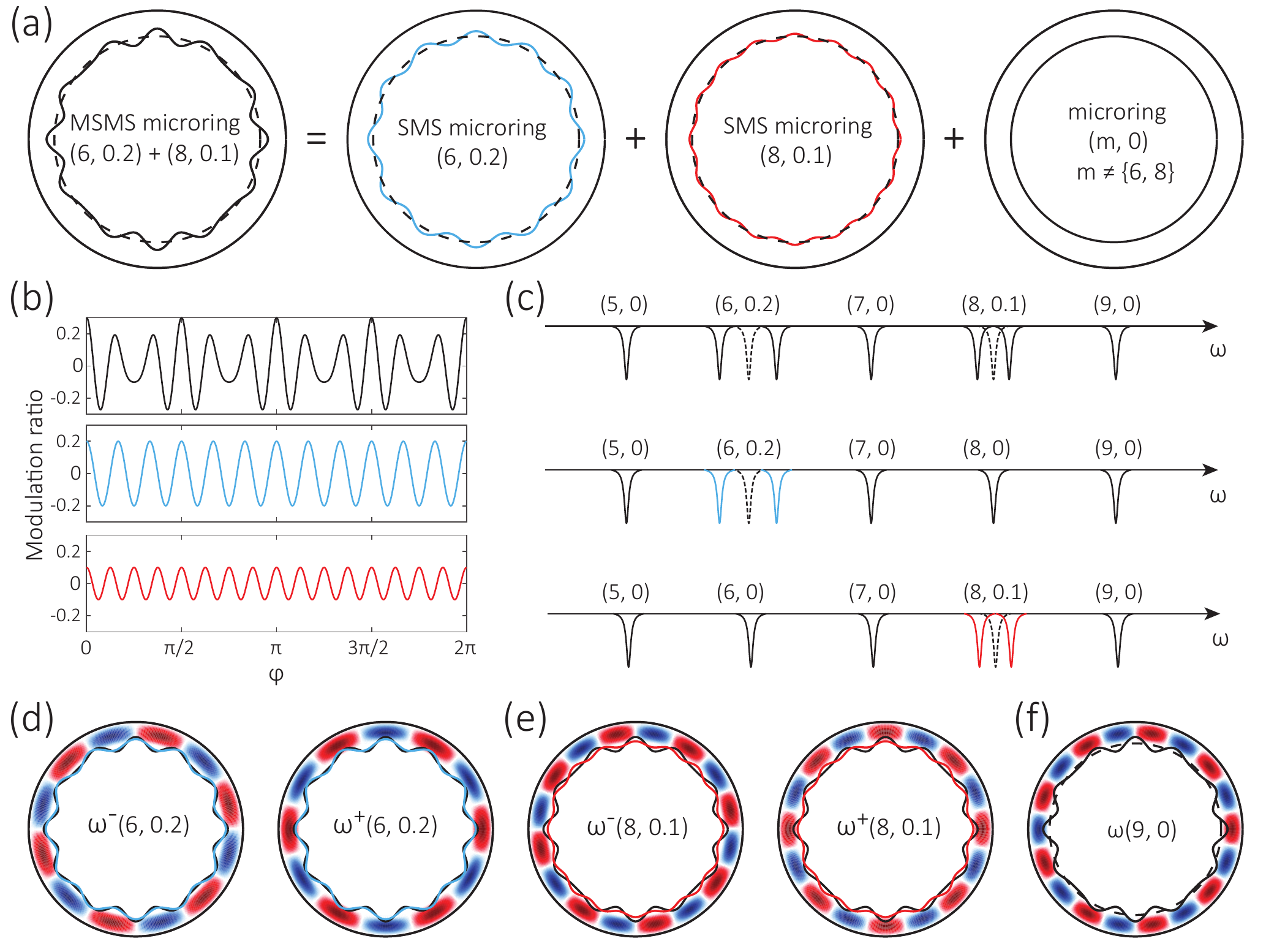}
\caption{\textbf{Illustration of multiple selective mode splitting (MSMS).} \textbf{(a)} Example of microring ring width modulation targeting $m$ = 6 and $m$ = 8 modes, with 20~\% and 10~\% of the nominal value (dashed line), respectively. This modulation selectively frequency splits the $m$ = 6 (d) and $m$ = 8 (e) modes, while behaving as a normal micoring with rotational symmetry for all other modes. We note that the $m$ number and modulation amplitude are quite different than that implemented in real devices, for illustration purposes. \textbf{(b)} The modulation of the ring width for the MSMS device is plotted versus azimuthal angle in the top panel, and has two frequency components in this case. The middle panel shows the modulation component targeting the $m$ = 6 mode with 20~\% modulation of the ring width, and the bottom panel shows the modulation component targeting the $m$ = 8 mode with 10~\% modulation of the ring width. \textbf{(c)} In the device transmission (top), this MSMS device will introduces frequency splitting of the $m$ = 6 and $m$ = 8 modes, and the amplitude of the splittings linearly depends on their modulation amplitudes. This transmission trace is equivalent to the two individual SMS transmission traces (middle and bottom) together. \textbf{(d)} For the $m$ = 6 mode, the MSMS device is equivalent to the SMS device described by the (6, 0.2) modulated boundary (blue). Degenerate clockwise and counter-clockwise modes are renormalized to two standing-wave modes. The standing-wave mode that always experiences a wider ring width has a larger resonance wavelength, and therefore a red-shifted resonance frequency. The other mode always experiences a narrower ring width and is blue-shifted. \textbf{(e)} For the $m$ = 8 mode, the MSMS device behaves as the SMS device described by the (8, 0.1) boundary (red). \textbf{(f)} For other modes ($m$ $\neq$ \{6,8\}), for example, $m$ = 9, the MSMS device does not induce coherent scattering, and therefore behaves similar to a rotationally-symmetric ring (dashed line).}
\label{Fig1}
\end{figure*}

WGM microcavities have been extensively studied for parametric nonlinear optics in a variety of material platforms including, but not limited to, silicon~\cite{Leuthold_NatPhoton_2010}, silicon nitride~\cite{Okawachi_OL_2011, Levy_OE_2011, Ferdous_NatPhoton_2011, Lu_NatPhys_2019, Li_NatPhoton_2016, Li_Optica_2017, Karpov_NatCommun_2018, Samara_OE_2019, Singh_Optica_2019, Li_PRA_2019, Lu_Optica_2019, Lu_SHG_arXiv_2020}, silicon carbide~\cite{Lukin_NatPhoton_2019, Guidry_arXiv_2020}, silicon dioxide~\cite{Vahala_NatPhys_2007, Fujii_OL_2017, Chen_PRL_2019}, aluminum nitride~\cite{Guo_Optica_2016, Bruch_Optica_2019, Tang_OL_2020}, gallium arsenide~\cite{ Kuo_NatCommun_2014,Lin_APLPhoton_2019}, magnesium fluoride~\cite{Liang_Optica_2015, Sayson_NatPhoton_2019, Fujii_OL_2019}, and lithium niobate~\cite{Ilchenko_PRL_2004, Lin_PRAppl_2016, JLu_Optica_2019, Luo_OE_2017, Zhang_Nature_2019}. Based on these platforms, many nonlinear processes have been demonstrated, including second harmonic generation~\cite{Levy_OE_2011, Guo_Optica_2016, Lin_PRAppl_2016, Luo_OE_2017, JLu_Optica_2019, Lukin_NatPhoton_2019, Lu_SHG_arXiv_2020},  third harmonic generation~\cite{Vahala_NatPhys_2007, Levy_OE_2011, Chen_PRL_2019}, optical parametric oscillation~\cite{Vahala_PRL_2004, Bruch_Optica_2019, Tang_OL_2020, Sayson_NatPhoton_2019, Lu_Optica_2019, Fujii_OL_2019, Guidry_arXiv_2020}, four-wave mixing Bragg scattering~\cite{Li_NatPhoton_2016}, cascaded four-wave mixing~\cite{Liang_Optica_2015, Fujii_OL_2017}, and electro-optic~\cite{Zhang_Nature_2019} and dissipative-Kerr-soliton~\cite{Ferdous_NatPhoton_2011, Okawachi_OL_2011, Li_Optica_2017, Karpov_NatCommun_2018} frequency comb generation. To match the mode frequencies in these processes, conventional methods demand accurate control of the device geometry (typically the microcavity cross-section) to achieve a specific global dispersion profile. Put in other words, all WGMs are shifted together but in a non-uniform way to achieve frequency matching. Though such conventional methods are particularly suitable for the case of frequency combs, where hundreds of WGMs participate in the nonlinear process, in many other cases the lack of individual mode control is a disadvantage. In particular, there are processes ranging from four-wave mixing Bragg scattering~\cite{Li_NatPhoton_2016} to widely-separated optical parametric oscillation~\cite{Lu_Optica_2019} where frequency matching of several (four and three, respectively) targeted modes should be accompanied by frequency mismatching of potentially competing processes. Conventional methods, that is, adjusting the global dispersion profile through the engineering of the resonator cross-section, along with post-fabrication trimming~\cite{Lu_Optica_2019} and thermal tuning~\cite{Guo_Optica_2016}, may be inadequate in those situations.

In contrast, direct control of targeted WGM frequencies, without influencing other WGMs, would be a desirable tool for frequency engineering in nanophotonics. One approach towards this end is to couple the clockwise (CW) and counterclockwise (CCW) propagating modes of the resonator and split the cavity resonances. Such mode interactions have often been introduced by random scattering (e.g., by sidewall roughness) in a microcavity~\cite{Weiss_OL_1995}; however, this random scattering is difficult to control. Later on, the mode interaction introduced by a foreign point scatter to a microcavity was studied~\cite{Mazzei_PRL_2007}, and is particularly useful for particle sensing~\cite{Zhu_NatPhoton_2009}. More controllable frequency splitting has been induced using interacting degenerate modes from \xl{two}~\cite{Gentry_OL_2014} \xl{or multiple}~\cite{Smith_JOSAB_2003} adjacent microcavities. Within one microcavity, the coupling between CW and CCW modes can be realized by integrated distributed Bragg grating reflectors~\cite{Arbabi_APL_2011}, integrated phase shifters~\cite{Li_OL_2017}, and coherent modulation of the boundary, termed selective mode splitting (SMS)~\cite{Lu_APL_2014}. In particular, SMS allows accurate frequency control of a selected cavity mode, and has been used to help frequency matching in spontaneous four-wave mixing~\cite{Lu_APL_2014}, identify the azimuthal mode numbers of WGMs~\cite{Lu_NatPhys_2019}, and offset nonlinear frequency shifts in microcomb generation~\cite{Yu_arXiv_2020}. Here, we show that its potential for frequency engineering is much greater -- it can be controllably applied to multiple modes simultaneously, while leaving untargeted modes nearly unaffected. Such control is of high value to frequency mixing in nonlinear optics, where multiple modes naturally interact.

We introduce our approach, termed multiple selective mode splitting (MSMS), in the following sections. We first develop a basic model of MSMS using perturbation theory (Section~2). We then examine this idea through experiments (Section~3) that explore the extent to which we can selectively target specified WGMs with a controllable frequency splitting while minimally influencing other (untargeted/collateral) modes. We find that the splitting can be controlled by the modulation amplitude in a linear fashion, and that the collateral splitting is over $13~\times$ smaller, even for immediately adjacent WGMs. We then demonstrate several device configurations in which five different WGMs are simultaneously targeted and individually controlled, all while retaining cavity $Q>10^5$. We find that MSMS modes have approximately half the waveguide-resonator coupling rate of untargeted modes, consistent with their expected standing-wave nature (Section~4). Finally, we discuss scenarios in which MSMS can be applied to frequency match nonlinear processes, thereby highlighting the potential broad usage of this technique (Section~5). Although our experiments are implemented in the silicon nitride platform with four-wave mixing processes, the
MSMS method is independent of underlying material platform and the specific resonator global dispersion, and can be applied to both $\chi^{(2)}$ and $\chi^{(3)}$ processes. We therefore believe that MSMS can be a powerful and universal tool to aid frequency engineering in microcavity nonlinear optics.

\section{Theory}
A rotationally-symmetric (i.e., without azimuthal modulation) microring resonator supports cavity modes propagating in opposite directions, clockwise (CW) and counter-clockwise (CCW), which are degenerate in frequency $\omega^\text{0}_m$, where $m$ is the azimuthal mode number. A modulation of the microring geometry, for example, of its ring width as shown in Fig.~\ref{Fig1}(a), with an angular frequency ($n$ oscillations within the ring circumference) matching that of the targeted optical mode $n = 2m$ (Fig.~\ref{Fig1}(d,e)) introduces coupling, at a rate of $\beta_m$, between the CW and CCW modes. This coupling lifts their frequency degeneracy and results in standing-wave modes at two new frequencies, $\omega^{\pm}_m = \omega^0_m \pm \beta_m$, each containing CW and CCW contributions~\cite{Borselli_OE_2005}. The standing-wave mode experiencing a smaller (larger) effective ring width has a shorter (longer) resonance wavelength, i.e., a higher (lower) resonance frequency, as shown in Fig.~\ref{Fig1}(d,e). Their frequency difference, i.e., the magnitude of the mode splitting $2\beta_m$, depends on the amplitude of the geometric modulation applied at that specified angular frequency ($A_{n=2m}$), as shown in Fig.~\ref{Fig1}(b,c).
Therefore, by changing this modulation amplitude with sufficient resolution, we can precisely control the cavity frequencies. Moreover, for the case $n \ne 2m$, as shown in Fig.~\ref{Fig1}(f), because the modulation is not in phase with optical mode, the optical mode should remain unperturbed (except $n = 0$, which corresponds to a uniform change of ring width). This orthonormal property is at the heart of MSMS, as schematically depicted in Fig.~\ref{Fig1}(a), and can be deduced from a perturbation theory of optical modes with shifting boundaries~\cite{Johnson_PRE_2002},
\begin{eqnarray}
\beta_m = \frac{\omega_m}{2} \frac{\int{d S \cdot A \left[ (\epsilon_\text{c} - 1)(|E_{\parallel}|^2) + (1 - 1/\epsilon_\text{c})|D_{\perp}|^2\right]}}{\int d V {\epsilon(|E_{\parallel}|^2 + |E_{\perp}|^2)} }, \quad \label{eq1}
\end{eqnarray}
where $E_{\parallel}$ ($D_{\parallel}$) and $E_{\perp}$ ($D_{\perp}$) are the electric field components (displacement field components) of the unperturbed optical mode that are parallel ($\parallel$) and perpendicular ($\perp$) to the modulation boundary $d S$, respectively. $\epsilon$ represents the dielectric function of the material, including the silicon nitride core ($\epsilon=\epsilon_\text{c}$), silicon dioxide substrate, and air cladding ($\epsilon=1$).

For transverse-electric-field-like (TE) modes or transverse-magnetic-field-like (TM) modes, the above equation can be further simplified since the term containing either $D_{\perp}$ or $E_{\parallel}$ dominates the integral in the numerator. Take TE-like modes for example, with the proposed boundary modulation on the ring width,
\begin{eqnarray}
W(\phi) = W_\text{0} + \sum_{n}{A_n \text{cos}(n\phi)},  \quad \label{eq2}
\end{eqnarray}
The standing-wave mode with a larger frequency has a dominant displacement field $D(r,\phi,z)=D(r,z)\text{cos}(m \phi)$. Equation~(\ref{eq1}) can be written as:
\begin{eqnarray}
\beta_m = \sum_{n} \frac{A_n \omega_m}{2} \frac{ \int{d S (1 - 1/\epsilon_\text{c})|D(r,z)|^2 \text{cos}^2(m \phi) \text{cos}(n \phi) }}{\int {d V \epsilon (|E_{z}|^2+|E_{\phi}|^2+|E_{r}|^2) \text{cos}^2(m \phi) } }. \quad \label{eq3}
\end{eqnarray}
The azimuthal part can thus be integrated separately,
 \begin{eqnarray}
\beta_m = \sum_{n} g_m A_n \int_0^{2\pi}{d \phi \text{cos}^2(m \phi) \text{cos}(n \phi)} / \pi = g_m (A_0 + \frac{A_{2m}}{2}), \quad \label{eq4}
\end{eqnarray}
where $g_m$ is defined as
\begin{eqnarray}
g_m \equiv \frac{\omega_m}{2} \frac{\int{d S (1 - 1/\epsilon_\text{c})|D(r,z)|^2}}{\int {d V \epsilon (|E_{z}|^2+|E_{\phi}|^2+|E_{r}|^2) }  }. \quad \label{eq5}
\end{eqnarray}
Importantly, Eq.~(\ref{eq4}) clearly shows that optical modes are only split when they are targeted in the modulation ($n = 2 m$). The coupling strength of a targeted mode scales linearly with modulation amplitude ($A_{2m}$), with a coupling coefficient ($g_m$) defined by Eq.~(\ref{eq5}). The value of $g_m$ can be evaluated from Eq.~(\ref{eq4}) by a uniform change of ring width ($A_0$).
Put in other words, the frequency shift induced by the sinusoidal modulation of $A_{2m} \text{cos}(2m\phi)$ is identical to the frequency shift of a uniform ring-width shift of $A_0=A_{2m}/2$. Therefore, in principle, one could split many optical modes with different amounts of mode splitting simply by setting a modulation of the boundary of the ring as in Eq.~(\ref{eq2}).

\section{Experiments}
\begin{figure*}[t!]
\centering\includegraphics[width=0.90\linewidth]{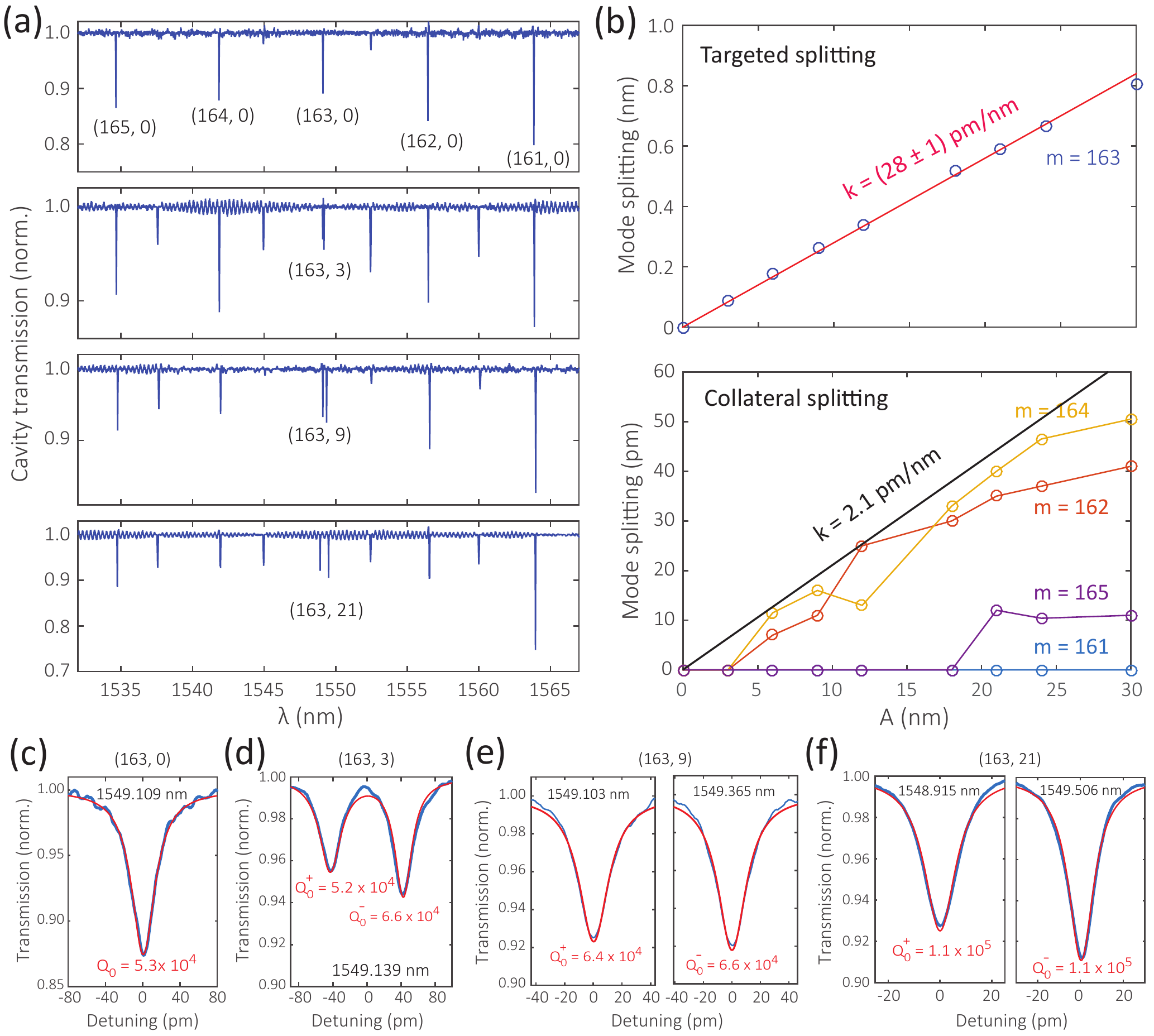}
\caption{\textbf{Single-mode selective mode splitting (1SMS).} \textbf{(a)} Transmission scans for four devices with different modulation amplitudes, where the labels ($m$, $A$) represent the azimuthal mode number $m$ and the (nominal) modulation amplitude $A$ in nanometers. In the top panel, the control device without SMS shows a cavity transmission spectrum without frequency splitting for any of the modes. From the second panel down, the devices exhibit a transmission spectrum with increasing frequency splitting for the targeted $m$=163 mode, with a splitting proportional to the prescribed modulation amplitude. \xl{Besides the
TE1 modes under investigation, another set of modes appears, which is verified to be fundamental transverse-magnetic modes (TM1).} The appearance of these modes is due to wavelength-dependent polarization rotation in the fibers used, which could be resolved by using polarization-maintaining fibers. \textbf{(b)} The top panel shows that the amount of mode splitting is linearly proportional to the modulation amplitude and reaches a value of $\approx$ 0.8 nm. The bottom panel shows that the adjacent four modes have much smaller mode splitting ($<$ 7.5~\% of the targeted mode, that is, below the black line) than the targeted modes. In both panels the uncertainties in the mode splitting, determined from nonlinear least squares fits to the data, are smaller than the data point size. \textbf{(c)} Transmission spectrum and fit trace of the singlet resonance without mode splitting, (163, 0), where $Q_\text{0}$ is the intrinsic optical quality factor. \textbf{(d)} Transmission spectrum for the doublet resonance (163, 3), with an observable mode splitting that is a couple of times the mode linewidths. From these linewidths we determine $Q^{+}_\text{0}$ and $Q^{-}_\text{0}$, the intrinsic optical quality factors for the blue-shifted and red-shifted modes, respectively. \xl{The differences in the coupling and intrinsic Qs for these two modes are potentially due to the different localization of the two standing-wave modes with respect to the ring-waveguide coupling region and possible point defects, respectively.} \textbf{(e-f)} Two doublet resonances (163, 9) and (163, 21) that are well separated (262~pm and 591 pm splitting, respectively). See Appendix C for fitting methods.}
\label{Fig2}
\end{figure*}
In the previous section, we showed how a simple theory predicts that MSMS should achieve our goals, namely, selective frequency splitting of targeted modes with a controllable splitting amplitude. Next, we examine how well this technique works in practice. Of particular interest is our ability to experimentally create a sufficiently faithful modulation pattern on the inner surface of the microring to realize the mode-selective, adjustable frequency splitting we desire. For this purpose, we use a fabrication process (see Appendix D) in the Si$_3$N$_4$/SiO$_2$ platform without high-temperature annealing and hydrofluoric-acid cleaning processes. Though such processes can reduce optical loss, they are known to influence fabricated dimensions (including smearing out small features by up to several nanometers), and for our purposes here we desire to keep the fabricated structures as close to the intended designs as possible.

\subsection{Single-mode SMS (1SMS)}
We first examine the control of the frequency splitting of a single azimuthal mode, which we term single-mode selective mode splitting (1SMS). Here, we target a fundamental TE mode with $m = 163$, which has resonance around 1550~nm in a device with nominal parameters of a 25~$\mu$m outside ring radius, a 1.1~$\mu$m ring width, and a 500~nm thickness. Such a structure supports only fundamental radial modes in both TE and TM polarization. When there is no modulation, as shown in the top panel of Fig.~\ref{Fig2}(a) and in Fig.~\ref{Fig2}(c), the cavity transmission consists only of singlet azimuthal modes without any mode splitting. In contrast, when every other parameter is fixed and the ring width is modulated by the function W($\phi$)= 3~nm~$\cdot$cos(2$\times$163$\cdot\phi$), the mode with $m=163$ splits into two resonances ($\approx$ 85 pm apart), as shown in Fig.~\ref{Fig2}(d), while all other modes are still singlet resonances without splitting (the second panel of Fig.~\ref{Fig2}(a)). A further increase of the modulation amplitude increases the splitting magnitude more, and the 9~nm and 21~nm modulation cases are shown in the third and fourth panels in Fig.~\ref{Fig2}(a), with zoom-in traces of the split modes shown in Fig.~\ref{Fig2}(e)-(f).

In Fig.~\ref{Fig2}(b), we analyze the dependence on modulation amplitude of the mode splitting for the targeted and un-targeted modes. Here, we display data from devices with modulation amplitudes varying from 0 nm to 30 nm with a 3 nm step, with the exception of two cases (15 nm and 27 nm modulation amplitudes) that were damaged during fabrication. In the top panel, we can see a clear linear dependence of the targeted mode splitting versus the modulation amplitude, where the splitting is $k~=~$(28~$\pm$~1)~pm per 1 nm modulation of the inside ring width, and the uncertainty is a one standard deviation value based on a fit of the data to a linear function. Importantly, such linear dependence holds true even for a 3~nm modulation amplitude, which is only slightly larger than the pattern resolution (1~nm). On the high amplitude side, at a modulation of 30 nm, the mode splitting approaches 0.8 nm and is not apparently saturated. The extent to which the linear dependence holds for modulation amplitudes $<$~3~nm and $>$~30~nm requires further examination.

Importantly, the collateral splitting, i.e., the splitting for un-targeted modes, is comparatively small. As shown in the bottom panel of Fig.~\ref{Fig2}(b), the four adjacent modes ($m = \{161, 162, 164, 165\}$) all exhibit $<$~7.5~\% of the splitting of the targeted mode ($m = 163$), indicated by the black line. While it is natural to understand the second-most adjacent modes ($\delta m$ = $\pm$ 2) are less perturbed than the most adjacent modes ( $\delta m$ = $\pm$ 1), it is unclear to us why the $\delta m$ = 2 modes are apparently more affected than $\delta m$ = -2 modes, as the $m = 161$ modes seem not affected at all. Overall, such small collateral splitting supports the theory that, to a significant extent (i.e., $<$ 7.5~\%), un-targeted modes are unaffected by the targeted modulation, which makes it possible to extend this mode splitting method to the regime of controlling multiple modes simultaneously. In particular, the collateral splitting for 3 nm modulation amplitude is near zero (unresolvable in the transmission measurement) for all the un-targeted modes, and the modes are well fit to singlet Lorentzians, while the targeted splitting is as large as $\approx$ 85 pm.

A natural question is whether the introduced modulation adversely effects the modal quality factors. Other works utilizing single mode-splitting~\cite{Lu_NatPhys_2019,Yu_arXiv_2020} have observed that high $Q>10^5$ can be maintained, and in this work we consistently observe $Q\sim10^5$ when large mode splittings are realized. In fact, to our surprise, with a larger mode splitting, the $Q$s are not decreased but are actually increased compared to the zero-splitting or small-splitting case (see Appendix A). We also observe an increase in $Q$ for devices employing MSMS versus SMS, which is discussed later in Section V. Finally, we note that the devices studied in this work have intrinsic $Q$s that are in general about 10 times smaller than the state of the art for Si$_3$N$_4$ microrings ($Q_\text{0} \approx 10^6$). We believe this is at least partly due to the conservative nanofabrication method used in this work (in particular, avoiding the high-temperature annealing step). \ks{In Appendix E, we present measurements of a device in which annealing and post-fabrication wet etching are used to improve the material/surface quality, resulting in $Q$s approaching 10$^6$ for mode splitting levels in excess of 1.6~nm. Further study is required to determine what fundamental limitations in $Q$ the adoption of SMS/MSMS imposes, and to what extent these post-fabrication processes can mitigate loss processes while retaining adequate pattern fidelity}.

\begin{figure*}[t!]
\centering\includegraphics[width=0.90\linewidth]{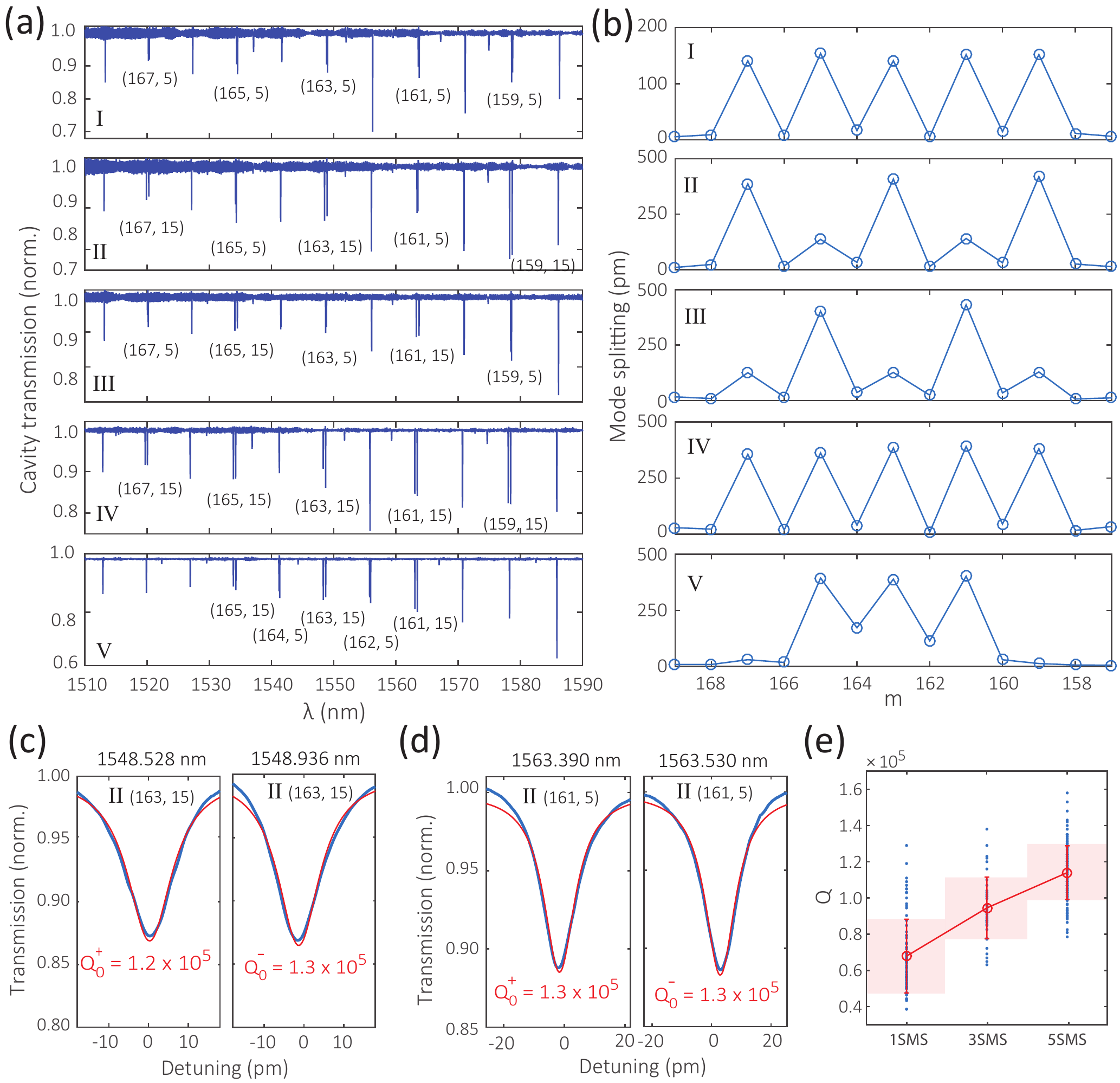}
\caption{\textbf{Five-mode selective mode splitting (5SMS).} \textbf{(a)} Cavity transmission traces of five devices with different configurations of five-mode selective mode splitting. Each split mode is labeled by ($m$, $A$), with $A$ units of nanometers. \textbf{(b)} The measured mode splitting amounts clearly correspond to the expected modulation pattern, where $A=15$~nm and $A=$5~nm ring width modulation amplitudes correspond to (390~$\pm$~20)~pm and (140~$\pm$~10)~pm mode splittings, while the un-targeted modes have $\approx$ 7~$\%$ collateral splitting. The slopes (splitting divided by the modulation amplitude) are (26~$\pm$~1)~pm/nm and (28~$\pm$~2) pm/nm, respectively, close to the 1SMS slope in Fig.~\ref{Fig2}(b). The uncertainties in mode splitting are smaller than the data point size. \textbf{(c)} Typical example of the cavity transmission for a targeted mode with 408 pm  mode splitting. \textbf{(d)} Typical example of the cavity transmission of a targeted mode with 140 pm mode splitting. \textbf{(e)} The average of the intrinsic optical quality factors of all the modes in these 5SMS devices is, surprisingly, more than one standard deviation higher than the average in the 1SMS devices (red open circles). $Q$s for individual modes are shown as solid blue dots. We note that the $Q$s of 1SMS modes are generally higher with $>$~20~nm modulation amplitude (see Appendix A). The 3SMS devices (where three modes are targeted for mode splitting) have $Q$s that are in-between the 1SMS and 5SMS cases (see Appendix B).}
\label{Fig3}
\end{figure*}

\subsection{Multiple SMS (MSMS)}
We now set up the modulation of the inner ring boundary to target multiple modes, considering both three modes (3SMS) and five modes (5SMS). As the results are qualitatively quite similar (other than the number of targeted modes), we focus on the 5SMS cases in the main text, with the 3SMS case considered in Appendix B. In both cases, the modulation is simply a superposition of sinusoids, with frequencies chosen in accordance with the targeted azimuthal modes and amplitudes that can be adjusted as desired.

For 5SMS, we consider five devices with different modulation configurations, and their measured transmission traces are shown in Fig.~\ref{Fig3}(a). The modulated modes are labeled by ($m$, $A$), where $m$ represents the mode number (from 159 to 167), and $A$ represents the modulation amplitude in nanometers (either 5 nm or 15 nm). The measured mode splittings are shown in Fig.~\ref{Fig3}(b), with the pattern matching the targeted configuration very well - only targeted modes show significant mode splitting and the splittings are independently adjustable (even for adjacent modes). For example, device I has an average collateral splitting of (9.6~$\pm$~4.2)~pm, which is $\approx$ 6.5~$\%$ of the targeted splitting (148~$\pm~$6)~pm; device IV has an average collateral splitting of (27~$\pm$~12)~pm, which is $\approx$ 7.2~$\%$ of the targeted splitting (375~$\pm$15)~pm. \ks{Though small, these collateral splittings indicate that the fabricated modulation pattern on the inner ring sidewall is imperfect, and contains undesired non-zero spatial Fourier components. Nevertheless,} this first demonstration of independent frequency control of multiple whispering gallery modes represents a significant advance in microcavity frequency engineering, beyond global resonator dispersion engineering based on changing the geometric cross-section~\cite{Moss_NatPhoton_2013} or techniques such as quasi-phase matching~\cite{Ilchenko_PRL_2004, JLu_Optica_2019}. In Section V, we discuss how this type of modal frequency control can be applied to different nonlinear optical processes.

Clearly, selective frequency control of multiple cavity modes is of most benefit if the cavity $Q$s remain high. Counter-intuitively, we have observed that the complicated modulation patterns that split five cavity modes do not decrease the $Q$ of the devices, but increase them instead. This trend is displayed in Fig.~\ref{Fig3}(e). Here, for the 3SMS and 5SMS devices, all targeted modes, as well as the two adjacent modes, are analyzed and the resulting mean $Q$ and one standard deviation spread in the $Q$ are shown. For the single-mode mode splitting case (1SMS), the displayed data is based on results from all modulation amplitudes, for which we note that a larger modulation amplitude tends to increase the optical $Q$ as well (see Appendix A). Importantly, these devices are fabricated on the same chip, so this trend is seemingly unlikely to be due to fabrication fluctuation. The fundamental physical reasoning behind this relative increase in $Q$ with increasing modulation amplitude and number of split modes requires further investigation, and may be specific to the $Q$s observed in this work (5~$\times$~10$^4$--1.5~$\times$~10$^5$), as opposed to being a more universal trend for any $Q$s. Nevertheless, we note that the optical $Q$s achieved ($\gtrsim$~$10^5$) for large mode splitting amplitudes (up to 430 pm) are already very promising for the usage of MSMS in nonlinear optics.

\subsection{Microring-waveguide Coupling}
\begin{figure}[t!]
\centering\includegraphics[width=1.0\linewidth]{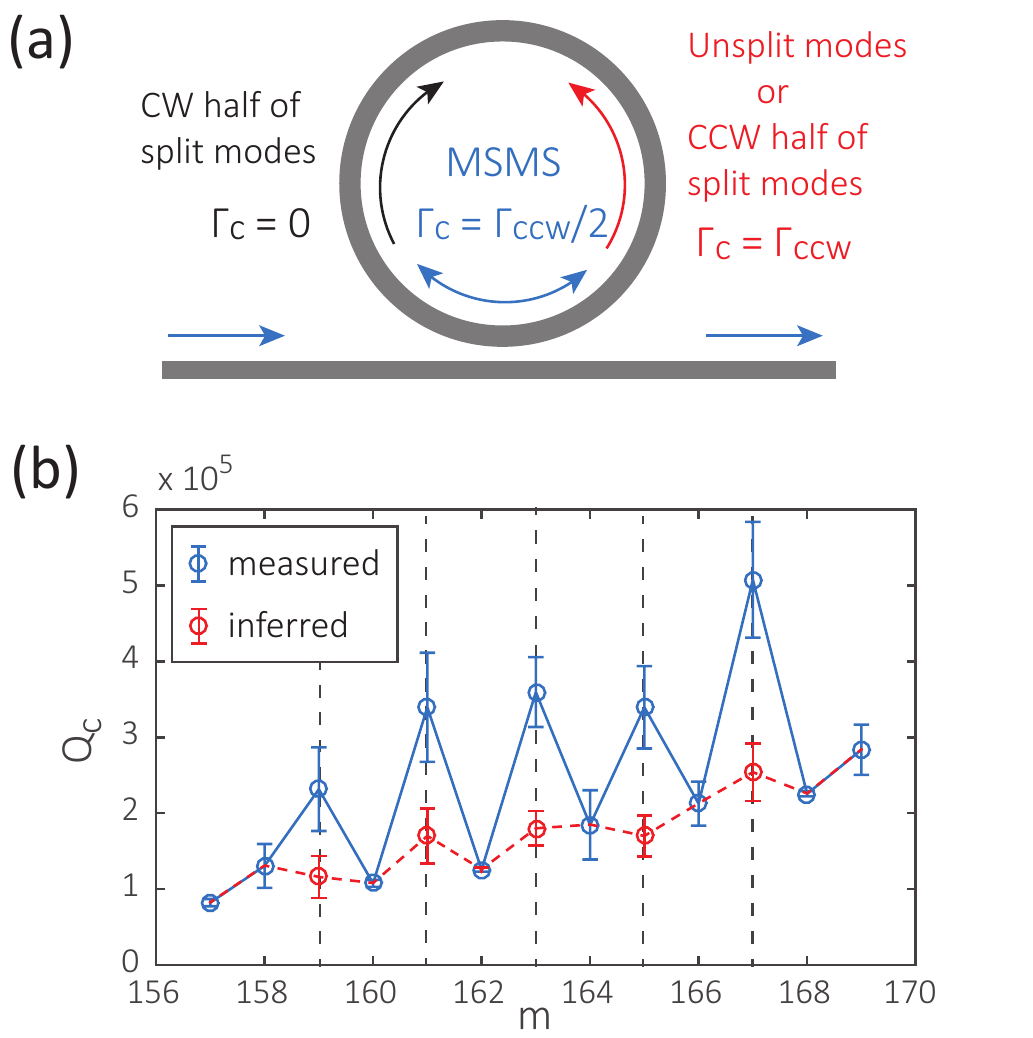}
\caption{\textbf{Microring-waveguide coupling.} \textbf{(a)} In a MSMS device, targeted modes become standing waves that are equally comprised of a counter-clockwise (CCW, red) part and a clockwise (CW, black) part. Assuming a undirectional waveguide input as shown in the diagram, the CCW part can be coupled to at a rate $\Gamma_\text{c} = \Gamma_\text{ccw}$, while the CW part is not directly coupled to from the waveguide because of the large momentum mismatch from the input light ($\Gamma_\text{c} = 0$). The MSMS modes are therefore expected to have an overall coupling rate of $\Gamma_\text{c} = \Gamma_\text{ccw}/2$ in theory. \textbf{(b)} Considering a 5SMS device in which the $m$=\{159,161,163,165,167\} modes are targeted, the measured $Q_{\text{c}}$ (blue) is clearly larger for the split modes (each indicated by a black dashed line) than other modes, and therefore $Q_{\text{c}}$ as a function of azimuthal mode order shows an oscillatory behavior. Dividing $Q_{\text{c}}$ for the split modes by 2 yields inferred values of their CCW coupling $Q$ (red), which are similar to those of the un-split modes. This trend is consistent with the prediction in (a), as $Q_\text{c}=\omega/\Gamma_\text{c}$. Error bars are one standard deviation values resulting from nonlinear least squares fits to the data.}
\label{Fig4}
\end{figure}

Aside from the change in intrinsic optical $Q$ introduced by MSMS, we also observe a change in the extrinsic (coupling) optical $Q$, which is related to the microring-waveguide coupling rate $\Gamma_\text{c}$ by $Q_\text{c} = \omega/\Gamma_\text{c}$. In theory, traveling-wave modes and standing-wave modes should have different coupling rates~\cite{Borselli_OE_2005}, as illustrated in Fig.~\ref{Fig4}(a). Light propagating in the forward direction from a waveguide underneath the microring couples to the microring CCW traveling-wave mode at a rate of $\Gamma_\text{c}$, while  the traveling-wave mode in the other direction, the CW mode, is uncoupled. In the MSMS case, the modes are standing waves, which are equally composed of CCW and CW components. Therefore, the forward propagating waveguide mode only directly excites the CCW part, leading to an overall reduction of the coupling rate to $\Gamma_\text{c}/2$, and therefore an increase in $Q_\text{c}$ by a factor of two relative to the traveling-wave case.

We indeed observe this coupling behavior in experiment. For the 5SMS device targeting azmiuthal mode orders $m$=$\{$159, 161, 163, 165, 167$\}$ with a 5 nm modulation amplitude each (top panel in Fig.~\ref{Fig3}(a,b)), the coupling $Q$s extracted from fits clearly show an increase relative to the untargeted modes, as shown in Fig.~\ref{Fig4}(b). Dividing the coupling $Q$s of the targeted modes by two (dashed lines) results in them having similar coupling $Q$s as the untargeted modes. There is an overall trend in coupling $Q$ for which larger $m$ modes (corresponding to smaller resonance wavelengths) generally exhibit worse coupling (larger $Q_\text{c}$). This is a consequence of the evanescent mode overlap between waveguide and resonator modes decreasing at shorter wavelengths.

\section{Application Scenarios}
\begin{figure*}[t!]
\centering\includegraphics[width=1.0\linewidth]{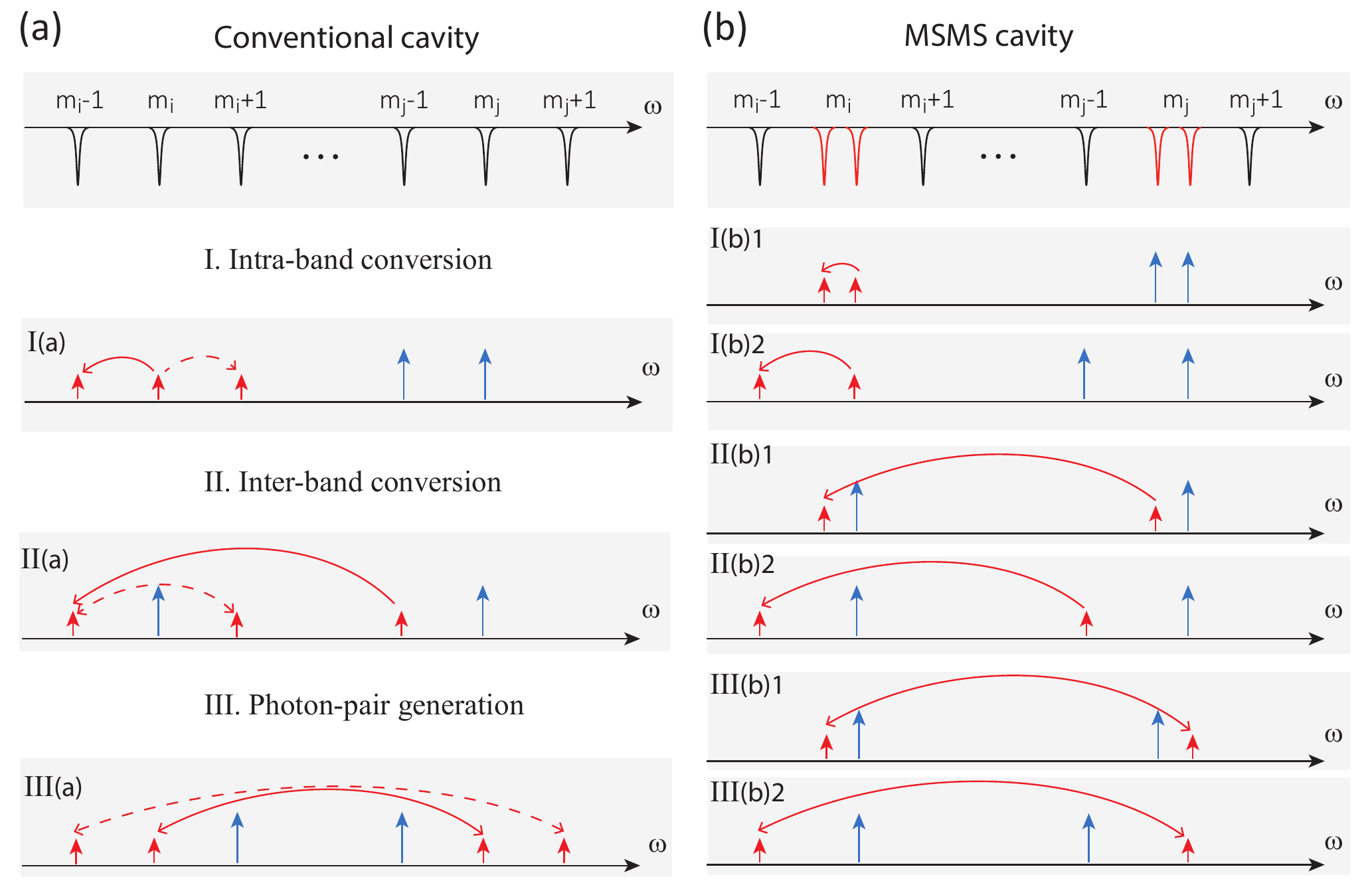}
\caption{\textbf{Nonlinear optics applications for the two-mode selective mode splitting (2SMS) case.} \textbf{(a)} In the case of a conventional WGM microcavity (left column), realizing an efficient parametric nonlinear process relies on finding a device geometry whose global dispersion profile results in frequency matching for the modes of interest for a given nonlinear optical process, e.g., intra-band (I) and inter-band (II) frequency conversion via four-wave mixing Bragg scattering, and photon-pair generation (III). \textbf{(b)} In contrast, by using a 2SMS device (right column), frequency matching can be achieved without any specific consideration of the global dispersion profile (and hence the resonator cross-section), so that any of the displayed nonlinear processes can be achieved. For intra-band FWM-BS in a conventional microcavity (I(a)), global dispersion engineering typically leads to an unwanted conversion channel (dashed arrow) along with the targeted channel (solid arrow). But in a MSMS cavity, FWM-BS naturally occurs for only a single set of modes (solid arrows in I(b)). Moreover, MSMS modes can be used in flexible ways, either exclusively MSMS modes (I(b)1), or combined with un-split modes (I(b)2). This MSMS cavity can also be applied to inter-band FWM-BS (II) and non-degenerately pumped photon pair generation (III) in similar ways, not only relaxing the frequency engineering for the targeted process (solid arrows), but also enabling suppression of the unwanted process (dashed arrows in I(a)--III(a)).}
\label{Fig5}
\end{figure*}

In previous sections, we have shown in the main text that MSMS can be a tool to split multiple cavity modes with a prescribed configuration of mode numbers and splitting amplitudes while retaining optical $Q~\gtrsim$~$10^5$. The induced modes are standing wave in nature and thus have a reduced microring-waveguide coupling rate relative to traveling wave modes. Although some aspects of the MSMS approach require further investigation and optimization, we believe that the demonstrated performance would already be of utility for many nonlinear optical applications. Here, we consider an example that showcases this capability. In comparison to a typical microring cavity (Fig.~\ref{Fig5}(a)), we use MSMS to frequency split two modes on demand ($m_\text{i}$, $m_\text{j}$) (Fig.~\ref{Fig5}(b)). Such on-demand two-mode splitting can be easily achieved based on the approach demonstrated in the previous section. We now discuss how this 2SMS microring can be used in the context of three four-wave mixing processes.

We first consider four-wave mixing Bragg scattering (FWM-BS), which is a theoretically noiseless process for frequency conversion down to the single-photon level~\cite{McKinstrie_OE_2005}. Recently, this process has been demonstrated in silicon nitride microrings~\cite{Li_NatPhoton_2016}, with intraband conversion (few nanometer spectral shifts) demonstrated for a quantum dot single-photon source~\cite{Singh_Optica_2019} and a spontaneous four-wave mixing source~\cite{Li_PRA_2019}. This intraband FWM-BS process involved two spectral bands that are widely separated from each other, with the pump modes, shown in blue in Fig.~\ref{Fig5}~(I(a)), situated in one band and the signal and idler modes (shown in red), situated in the other band. Frequency matching requires the signal-idler mode separation to equal the pump mode separation, and though achievable through global dispersion engineering that results in an equal free spectral range (FSR) in the two bands~\cite{Li_NatPhoton_2016}, there are limitations of that approach.  In particular, the wavelength shift is restricted to being a multiple of the FSR (modulo the cavity linewidth), and the smallest achievable shift is thus largely dictated by the ring radius. Moreover, frequency up- and down-shifts are realized simultaneously, as the equal FSR condition means that both idlers are equally well frequency matched. In contrast, the 2SMS case, shown on Fig.~\ref{Fig5}(I(b)), does not require any specific device dispersion or cross-sectional geometry. The achievable frequency shift is not limited to multiples of the resonator FSR, and very small frequency shifts can be achieved without going to very large resonator size. Moreover, in most cases, no other set of modes will be frequency matched, eliminating this competitive channel for frequency conversion, which is nontrivial to implement in a conventional design~\cite{Li_PRA_2019}. We note that this MSMS configuration can also be applied to inter-band frquency conversion based on FWM-BS (Fig.~\ref{Fig5}(II)), where in this case there are pump modes in each of the two spectral bands, and their frequency separation can be made to match the signal-idler mode frequency separation, so that large spectral shifts can be achieved~\cite{Li_NatPhoton_2016}.

Similarly, the same MSMS scheme can be used in photon pair generation. In previous work, for example, telecom-band pair sources~\cite{Samara_OE_2019} and visible-telecom pair sources~\cite{Lu_NatPhys_2019}, dispersion had to be tailored to phase- and frequency-match targeted modes (solid arrows), and unwanted photon pair channels (dashed arrows) need to be carefully suppressed, as shown in Fig.~\ref{Fig5}(III(a)). But in the MSMS case (Fig.~\ref{Fig5}(III(b))), the global dispersion profile is not important, and in fact, most geometries with nonzero dispersion will leads to noiseless photon pair generation, where all other mode configurations are not matched in frequency and phase. We note that, for specificity, we have shown a non-degenerate pumping configuration here, but there are many other configurations (including typical degenerate pumping) to which MSMS can be applied. A full discussion of such configurations would be valuable but is beyond the scope of this paper.

Finally, we note that the standing wave nature of the MSMS modes must be carefully considered in the context of the applications being discussed. These standing waves are intrinsically a superposition state of CW and CCW modes, a property that could be useful for helping in the generation of orbital angular momentum in single-photon sources~\cite{Jin_OAM_2020} and for enabling CW/CCW path entanglement~\cite{Rogers_CommunPhys_2019}. On the other hand, in many cases we want the output signal to exit the cavity in one direction, which will not be the case for standing wave modes. However, we note that the MSMS approach can still be valuable in such situations. In particular, frequency splitting may only be applied to pump modes, or MSMS may be used exclusively to frequency mismatch unwanted processes, while the process of interest is still accomplished using the typical traveling wave modes.

\section{Discussions}
In summary, we demonstrate a nanophotonic frequency engineering tool, multiple selective mode splitting (MSMS), which manipulates mutliple targeted cavity mode frequencies with individually controlled amounts of splitting. Such mode shifts can be large in amplitude (up to $\approx$ 0.8 nm), small in cross-talk (less than 7.5~\% for untargeted modes), and high-$Q$ ($\approx$ $10^5$). The MSMS tool enables unique frequency engineering capabilities beyond traditional techniques, can be universally applied regardless of materials and geometries, and will have broad applications in microcavity nonlinear photonics.

\section*{Appendix A: SMS Q analysis}
\begin{center}
\begin{figure}[t!]
\begin{center}
\includegraphics[width=1.0\linewidth]{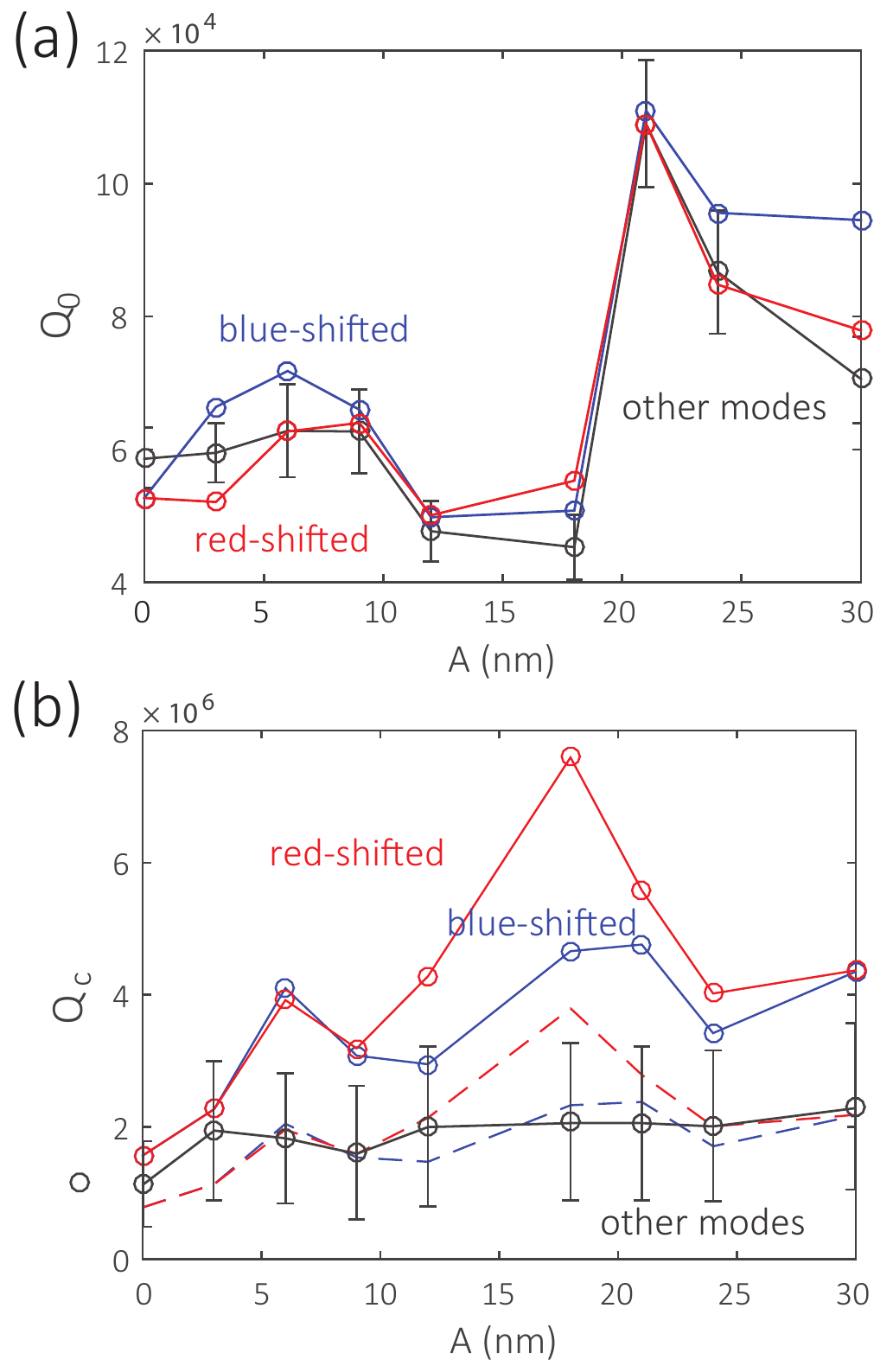}
  \caption{\textbf{1SMS $Q$ Analysis} \textbf{(a)} The intrinsic optical $Q$s of the SMS modes (red and blue) are consistent with the values from un-targeted modes (grey data indicates one-standard-deviation range given by fits to the four adjacent resonances). \textbf{(b)} The coupling $Q$s of the SMS modes (red and blue) are statistically larger than the values from un-targeted modes (one-standard-deviation range found by fits of the four adjacent resonances shown in grey), as expected based on their standing-wave nature. Dividing the coupling $Q$ of the targeted modes by two gives values (indicated by the red and blue dashed line) that are within the range of $Q$ values from other modes, consistent with expectation from theory.}
\label{FigS1}
\end{center}
\end{figure}
\end{center}

In this Section, we show additional data on the intrinsic and coupling Q analysis for the single-mode SMS devices. In Fig.~\ref{FigS1}(a), we see that the SMS modes (shown in red and blue) have similar intrinsic optical $Q$s as the un-targeted modes (shown in grey). Moreover, surprisingly, larger modulation amplitudes have higher intrinsic $Q$s than non-modulated devices. In particular, for a modulation amplitude of 21~nm, the modes have $Q>10^5$. However, this trend requires further verification through additional device fabrication and characterization in future.

The coupling $Q$ shows a statistical difference for SMS modes and un-targeted modes. In Fig.~\ref{FigS1}(b), we show that coupling Q of the SMS modes are in general larger (by about a factor of 2) than those values from the un-targeted modes, which agrees with the results in Section 4 in the main text.

\section*{Appendix B: Additional three-mode data}
\begin{center}
\begin{figure*}[t!]
\begin{center}
\includegraphics[width=1.0\linewidth]{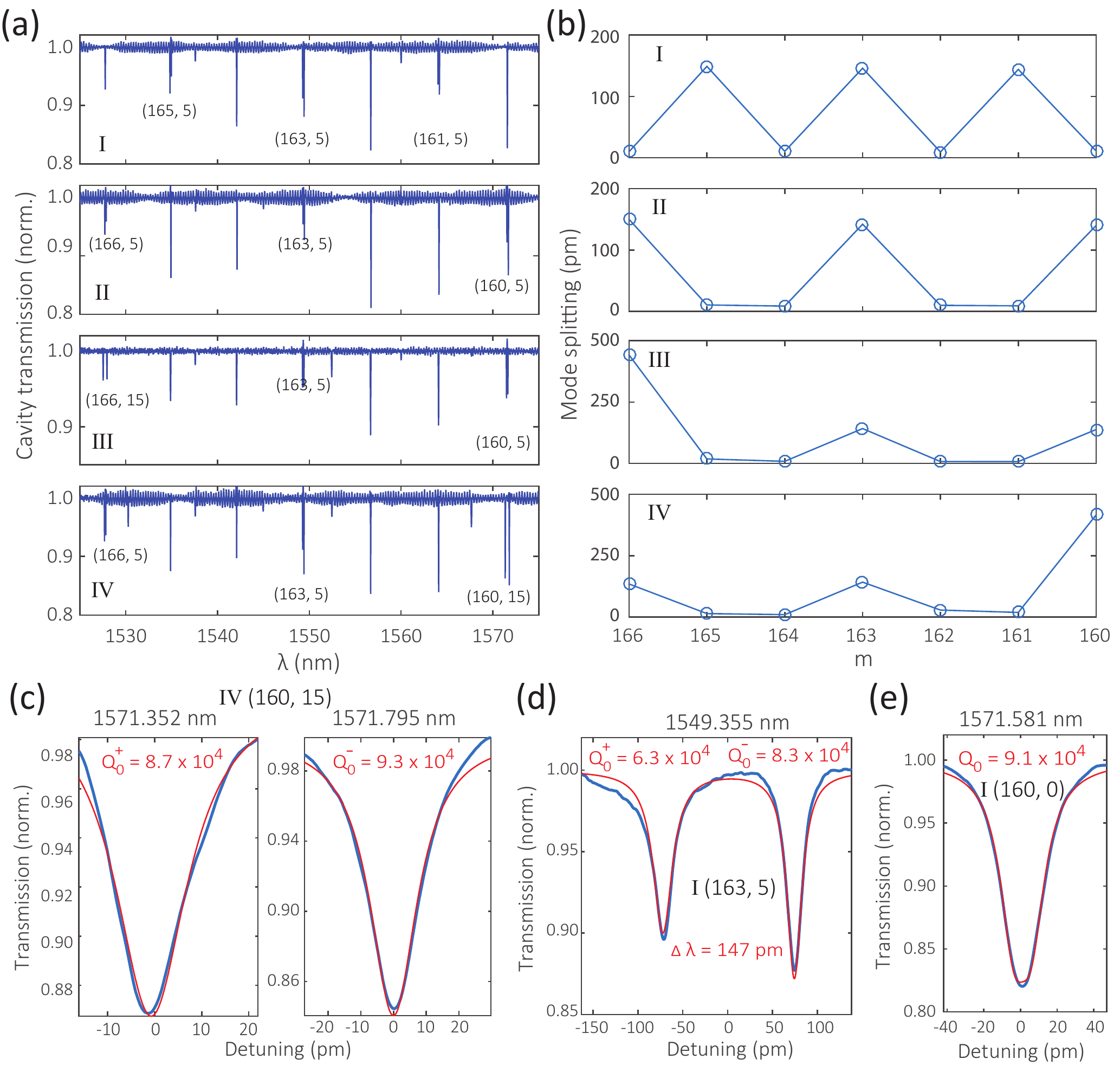}
  \caption{\textbf{Three-mode selective mode splitting results.} \textbf{(a)} Cavity transmission for four devices with different configuration of three-mode selective mode splitting. The modulated modes are labeled as ($m$, $A$), where $m$ represents the azimuthal mode number, and $A$ represents its modulation amplitude in nanometers. \textbf{(b)} The measured mode splittings clearly show the targeted control of three modes with a splitting amplitude proportional to the modulation amplitude. \textbf{(c-e)} Optical $Q$ fitting examples for modulation amplitudes of 15~nm, 5~nm, and 0~nm, respectively.}
\label{FigS2}
\end{center}
\end{figure*}
\end{center}

In the main text, we presented the compiled $Q$ data from 3SMS devices in Fig.~3(e), to examine the trend in intrinsic $Q$ vs. number of split modes. Here we show extended data from the 3SMS devices, where we target different WGMs with varying splitting levels. The results, shown in Fig.~\ref{FigS2}, are similar to the 5SMS devices.

\section*{Appendix C: Fitting method}
We explain our fitting method for singlet and doublet cavity resonances, following the method in Ref.~[\citen{Borselli_OE_2005}]. In general, there are three cases in Q fitting. First, the resonance is a singlet resonance without an observable splitting (e.g., Fig.~\ref{FigS2}(e)). Second, the resonance is a doublet resonance (e.g., Fig.~\ref{FigS2}(d)). Third, the resonance has a large mode splitting, that each mode looks like a singlet resonance (e.g., Fig.~\ref{FigS2}(c)). In the first case, we fit the resonance by a Lorentzian function, whose full-width at half-maximum yields the loaded optical quality ($Q_\text{t}$), and its transmission depth yields the ratio of $Q_\text{0}$ and $Q_\text{c}$. In this work, cavity modes are all under-coupled, therefore $Q_\text{0} < Q_\text{c}$. In the second case, we fit the resonances as two standing-wave modes that share the same coupling rate with three parameters, $\{ Q^\text{+}_\text{0}$, $Q^\text{-}_\text{0}$, $Q_\text{c} \}$. In the third case, we fit each standing-wave mode separately as an individual singlet, and allows these two modes to have different coupling Q values (i.e., fitting $\{ Q^\text{+}_\text{0}$, $Q^\text{+}_\text{c} \}$ and $\{ Q^\text{-}_\text{0}$, $Q^\text{-}_\text{c} \}$ respectively).

\begin{center}
\begin{figure*}[t!]
\begin{center}
\includegraphics[width=1.0\linewidth]{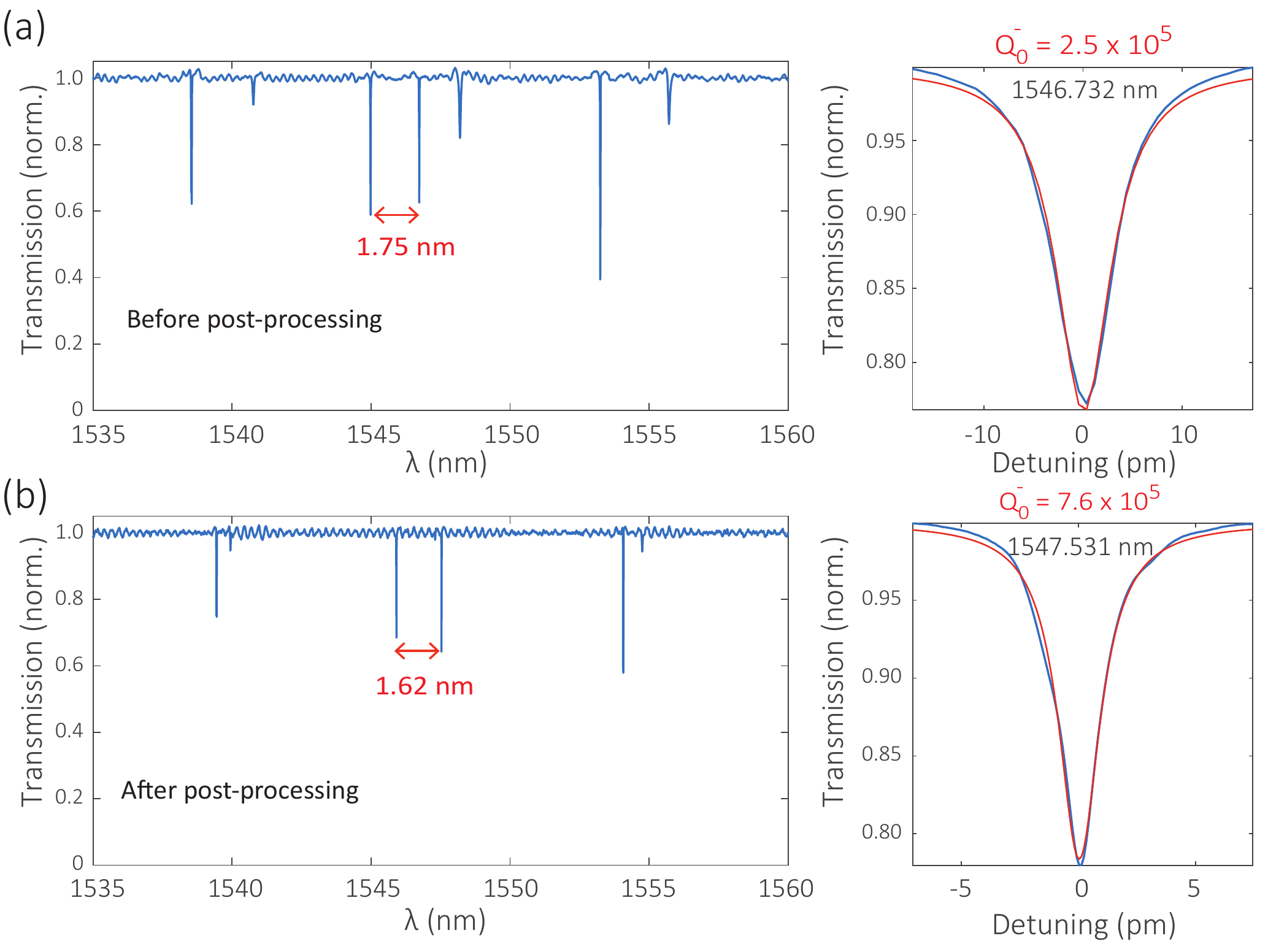}
  \caption{\textbf{High-Q mode splitting devices.} \textbf{(a)} An optimized etching process leads to a device with intrinsic $Q$ of $2.5 \times 10^5$  for a mode with a 1.75~nm splitting. \textbf{(b)} With post-processing steps consisting of annealing and dilute hydrofluoric acid etching, the device has shifted resonances with intrinsic $Q$ improved to $7.6 \times 10^5$ and decreased splitting of 1.62~nm.}
\label{FigS3}
\end{center}
\end{figure*}
\end{center}

\section*{Appendix D: Fabrication method} The device layout was done with the Nanolithography Toolbox, a free software package developed by the NIST Center for Nanoscale Science and Technology~\cite{coimbatore_balram_nanolithography_2016}. The ${\rm Si_3N_4}$ layer is deposited by low-pressure chemical vapor deposition on top of a 3~${\rm \mu}$m thick thermal ${\rm SiO_2}$ layer on a 100~mm diameter Si wafer. The wavelength-dependent refractive index and the thickness of the layers are measured using a spectroscopic ellipsometer, with the data fit to an extended Sellmeier model. Next, the device pattern is created in positive-tone resist by electron-beam lithography. The pattern is then transferred to ${\rm Si_3N_4}$ by reactive ion etching using a ${\rm CF_4/CHF_3}$ chemistry. The device is chemically cleaned to remove deposited polymer and remnant resist. An oxide lift-off process is performed so that the microrings have an air cladding on top while the input/output edge-coupler waveguides have ${\rm SiO_2}$ on top to form more symmetric modes for coupling to optical fibers. The facets of the chip are then polished for lensed-fiber coupling.

\section*{Appendix E: Higher Q devices}

\xl{In Fig.~\ref{FigS3}, we show transmission data of a device before and after post-processing (consisting of annealing and dilute hydrofluoric acid etching), to demonstrate that a large mode-splitting device can have optical quality factors approaching 10$^6$. Although its effect on the pattern accuracy needs further study, high-$Q$ modes approaching $7.6 \times 10^5$ are possible with $\approx$ 1.62 nm splitting. Such high-$Q$ and controllable mode splitting is very promising for nonlinear optical applications.}

\medskip
\noindent \textbf{Funding.}
This work is supported by the DARPA DODOS and NIST-on-a-chip programs.

\smallskip
\noindent \textbf{Acknowledgements.}
X.L. acknowledges support under the Cooperative Research Agreement between the University of Maryland and NIST-PML, Award no. 70NANB10H193.




\end{document}